\documentclass[journal]{IEEEtran}
\usepackage[numbers]{natbib}
\usepackage{algorithm}
\usepackage{algorithmic}
\usepackage{subfigure}
\usepackage{multirow}
\usepackage{multicol}
\usepackage{amsmath}
\usepackage{caption}
\usepackage{amsmath}
\usepackage{bm}
\usepackage{verbatim}
\usepackage{booktabs}
\usepackage{algorithm} 
\usepackage{algorithmic} 
\usepackage{multirow} 
\usepackage{xcolor}
\usepackage{caption}
\usepackage{graphicx}
\usepackage{subfigure}
\usepackage{float}
\usepackage{multirow}
\usepackage{multicol} 
\usepackage{colortbl,booktabs}
\ifCLASSINFOpdf
\else
\fi
\begin{document}
%
\title{Adaptive Multi-modal Fusion Hashing via Hadamard Matrix}
%
%
%

\author{Jun Yu,Donglin Zhang,Zhenqiu Shu,Feng Chen
\thanks{J. Yu is with the College of Computer and Communication Engineering,  Zhengzhou University of Light Industry, Zhengzhou, China.}
\thanks{D. Zhang is with the Jiangsu Provincial Engineering Laboratory
of Pattern Recognition and Computational Intelligence, Jiangnan University,
214122, Wuxi, China.}
\thanks{Z. Shu is with the Faculty of Information Engineering and Automation, Kunming University of Science and Technology, Kunming, China.}%
\thanks{F. Chen is with the School of Computer Science and Technology, Anhui University of Technology, Ma'anshan, China.}%
\thanks{(e-mail:yujunjason@aliyun.com)}}

%
%

\markboth{Journal}%
{Shell \MakeLowercase{\textit{et al.}}: Bare Demo of IEEEtran.cls for IEEE Journals}
%



\maketitle

\begin{abstract}
Hashing plays an important role in information retrieval, due to its low storage and high speed of processing. Among the techniques available in the literature,  multi-modal hashing, which can encode heterogeneous multi-modal features into compact hash codes, has received particular attention. Most of the existing multi-modal hashing methods adopt the fixed weighting factors to fuse multiple modalities for any query data, which cannot capture the variation of different queries. Besides, many methods introduce hyper-parameters to balance many regularization terms that make the optimization harder.  Meanwhile, it is time-consuming and labor-intensive to set proper parameter values. The limitations may significantly hinder their promotion in real applications. In this paper, we propose a simple, yet effective method that is inspired by the Hadamard matrix. The proposed method captures the multi-modal feature information in an adaptive manner and preserves the discriminative semantic information in the hash codes. Our framework is flexible and involves a very few hyper-parameters. Extensive experimental results show the method is effective and achieves superior performance compared to state-of-the-art algorithms. 
\end{abstract}


%
\IEEEpeerreviewmaketitle

\section{Introduction}
%
%
%
%
\IEEEPARstart{A}{s} an effective technique to deal with the challenges posed by the explosive growth of mutimedia data, hashing has attracted increasing attention in information retrieval and related areas \cite{wang2014hashing}. Existing hashing methods mainly focus on uni-modal hashing \cite{datar2004locality-sensitive,gong2013iterative,ji2017toward,lin2018supervised,shen2015supervised}.
Different from  uni-modal hashing where only a single modality is given as the research object, multi-modal hashing combines multiple modalities to comprehensively represent query data for multimedia retrieval. A simple way to extend uni-modal hashing to the multi-modal situation is to use a  representation concatenating  multiple uni-modal features  forming  the input of uni-modal hashing methods. However, such an extension may fail to exploit the complementarity of the modalities. To handle this problem,  various learning methods \cite{liu2012compact,Shen2015Multiview,Song2013Effective,Shen2018Multiview}  have been developed. Multiple Feature Hashing (MFH) \cite{Song2013Effective} explores the local structure of the individual features and fuses them in a joint framework. Multiple Kernel Learning (MKH) \cite{liu2012compact}  fuses multiple features by  an optimised  linear-combination. Multi-view Latent Hashing (MVLH) \cite{Shen2015Multiview} aims to find a unified kernel feature space where the weights of different modality are adaptively learned. Multiview discrete hashing (MVDH) \cite{Shen2018Multiview} jointly performs a matrix factorization and spectral clustering to learn compact hash codes. In MVDH,  the weight imposed on each modality is adaptively learned to reflect the importance of the modality to the learning process. Besides,  Some online adaptive hashing methods \cite{Lu2019Online,Zhu2020Flexible} are proposed. An example is Online Multi-modal Hashing with Dynamic Query-adaption (OMH-DQ) \cite{Lu2019Online} where a parameter-free online mode  can adaptively  learn the  hash codes for the dynamic queries. Although these approaches have achieved promising performance in many applications,  they have the following two shortcomings: (1) Most of the above multi-modal hashing methods model each modal information by constructing a similarity graph, which will cost $\mathcal{O}(n^2)$ and bring significant optimization challenge. This high computational complexity is not scalable to large-scale multimedia retrieval problems. (2) Some approaches usually incorporate additional regularization terms to enhance the discriminative capability. The hyper-parameters introduced to balance the terms to obtain optimal performance will require disordinate amount of time to adjust, which makes the methods inapplicable in practice.  

\begin{figure}[t]
	\centering
		\includegraphics[width=.5\textwidth]{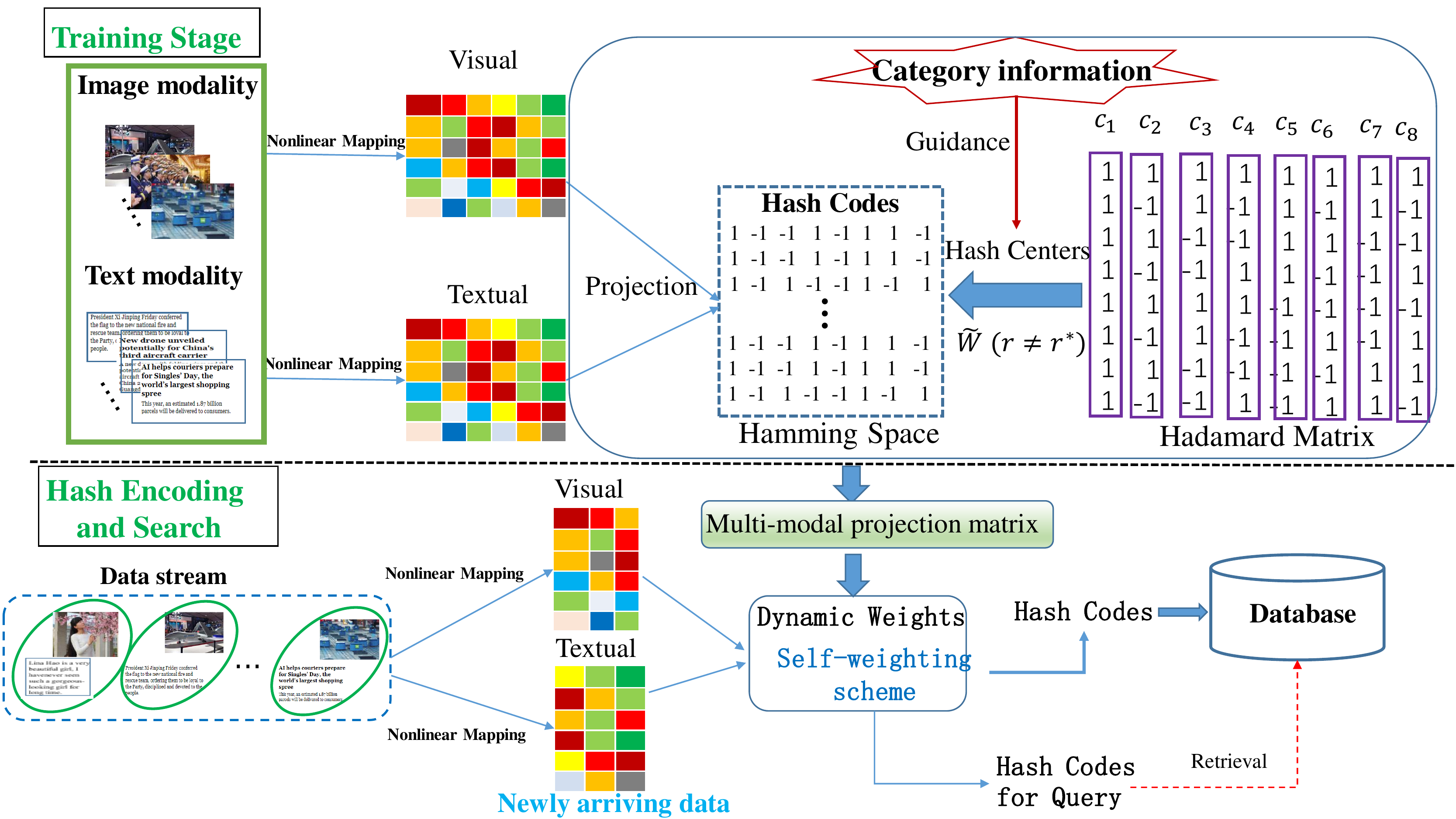}
	\caption{The overview of the proposed method. The proposed framework is divided into two parts: Training stage and Hash encoding stage. The multi-modal data is collaboratively projected into the common Hamming space where samples of the same category converge to 			their common hash center point generated by a Hadamard matrix  in the offline training stage. Based on the learned projection matrices in the offline training stage, we adopt an adaptive weighing scheme to obtain hash codes for new data in order to capture the 				variations of different samples in the hash encoding process.}
	\label{fig1}
\end{figure}

Based on the above observations, in this paper, we propose a simple yet very effective multi-modal hashing to overcome the challenging problem. Inspired by some recent works \cite{Koutaki2018Hadamard,Lin2020Hadamard,Yuan2020Central} where Hadamard matrix has been shown to be effective in the hash learning, we  introduce a Hadamard matrix to generate discriminative target codes for the data,  which induces the samples with the same label information to approach their common target codes in the hash function learning stage. In the hash encoding stage,  we adopt the adpatively self-weighting scheme to capture the dynamic information.  Figure \ref{fig1} illustrates the flowchart of the proposed framework. The advantages of our method are summarized as follows
\begin{itemize}
	\item We introduce a Hadamard matrix into the multi-modal retrieval process to guide the hash learning. We show that this enables discriminative semantic information to be preserved in the hash codes, although our model is relatively simple.
	\item The method is easy to implement and requires low computational time. It does not involve the setting of hyper-parameters.
	\item  A comparative evaluation of the proposed method with state-of-the-art hashing methods on three available \\datasets shows that our method boosts the retrieval performance.
\end{itemize}

Structurally,  the rest of this paper is organized as follows. In Section \ref{rw}, we review related works. The proposed model is described in Section \ref{method}. Section \ref{exeriments} analyses the experimental results and finally we draw the conclusions of the paper in Section \ref{conclusion}.

\section{Related Work}
\label{rw}
As mentioned previously, many hashing methods have been proposed to handle the multi-modal data, which can be classified into Cross-modal hashing and Multi-modal hashing. The cross-modal hashing\cite{hu2019collective,li2019deep,Liu2016Supervised,wang2018robust,yu2019discriminative,Donglin2020Learning} learns hash function for each modality to encode the hash codes of the corresponding modality. Multimodal Discriminative Binary Embedding (MDBE) \cite{wang2016multimodal} formulates the hash function learning in terms of classification and exploits the label information to discover the shared structures inside heterogeneous data. Lin Z. et al.\cite{lin2016cross} proposed Semantics-Preserving Hashing (SePH) that transforms semantic affinities into a probability distribution and approximates it with to be-learnt hash codes in Hamming space via minimizing the Kullback-Leibler divergence. Discrete Cross-modal Hashing (DCH) \cite{xu2017learning} is formulated to a joint optimization problem for learning the modality-specific hash functions and the unified binary codes. Discrete Latent Factor model based cross-modal Hashing (DLFH) \cite{jiang2019discrete} directly learns the binary hash codes without continuous relaxation. Enhanced Discrete Multi-modal Hashing (EDMH) \cite{chen2020enhanced} learns binary codes and hashing functions simultaneously from the pairwise similarity matrix of data. 

Different from cross-modal hashing, multi-modal hashing \cite{lu2020semantic,lu2019flexible,Lu2019Online,Shen2015Multiview,Song2013Effective,zheng2019fast,zheng2020efficient,Zhu2020Flexible}learns a unified hash function for paired multi-modal data. In other words, the paired modalities share the fused hashing features. Compared with cross-modal hashing, it can better represent the target object for out-of- sample extensions. Song J. et al. \cite{Song2013Effective} proposed  Multiple Feature Hashing (MFH) that preserves the local structural information of each individual feature and also globally considers the local structures for all the features to learn a group of hash functions. Multi-view Latent Hashing (MVLH) \cite{Shen2015Multiview} uses the latent factors shared by multiple views from an unified kernel feature space to learn the binary codes. Online Multi-modal Hashing with Dynamic Query-adaption (OMH-DQ) \cite{Lu2019Online} method is designed to adaptively preserve the multi-modal feature information into hash codes by exploiting their complementarity. Flexible Multi-modal Hashing (FMH) method \cite{Zhu2020Flexible} learns multiple modality-specific hash codes and multi-modal collaborative hash codes simultaneously, thus the combination of modality features are flexibly generated according to the newly coming queries. However, these multi-modal hashing approaches introduces additional parameters, whose optimal values need to be manually adjusted. This indicates that more time will be taken to adjust the parameters.  By the above analysis, it is very necesary to develop a simple yet effective method.

In this paper, we propose a novel multi-modal hashing algorithm that utilizes the advantage of the Hadamard matrix and the category information of data to guide the hash learning. In the hash encoding process,  the proposed model adopts the adaptive learning way to capture the dynamic difference of multi-modal data. The proposed method hardly need to set proper parameter manually in the entire learning process. 

\section{ THE PROPOSED METHOD}
\label{method}
\subsection{Model Formulation}\label{sec2}
Assume that the training dataset is comprised of $n$ multimedia instances with $M$ different modalities and category information. The $m$-th modality is denoted as $X^{(m)} = [x_1^{(m)},\\...,x_n^{(m)}]\in R^{d_m\times n}$, where $d_m$ is the dimensionality of the $m$-th modality.
Our method aims to learn the discriminative hash code $B \in \{-1,1\}^{r \times n}$ to represent multimedia instances, where $r$ is the length of the output codes in Hamming space.
We pre-define a set of points $C = \{c_1,c_2,...,c_k\}\\\in R^{r^* \times k}$ as the hash centers of specific-class respectively, where $k$ is the number of categories and $r^*(r^*= r)$ indicates the dimension of hash centers. We encourage data points with the same class information to be close to a common hash center and those with different semantic information to be associated with different hash centers respectively.
Intuitively, the pre-defined hash centers in the Hamming space should conform to the following requirement:  A sufficient mutual distance between the centers should be ensured so that samples from different classes are separated well in the Hamming space. The concept of  valid hash centers is specified in Definition 1\\
\textbf{Definition 1}. Hash centers $C = \{c_i\}_{i=1}^s\subset \{0,1\}^{r^*}$ in the $r^*$-dimensional Hamming space
satisfy that the average pairwise Hamming distance is greater than or equal to $r^*/2$, i.e.
\begin{equation}
\label{eq1}
\frac{1}{V}\sum_{i\neq j}^sD_H(c_i,c_j)\geq \frac{r^*}{2}
\end{equation}
where $V$ is the number of combinatins of different $c_i$ and $c_j$, $s$ is the size of set $C$, and $D_H$ denotes the Hamming distance.

Inspired by recent works \cite{Koutaki2018Hadamard,Lin2020Hadamard} which have been shown to be very promising in the field of hash learning, we introduce the Hadamard Maxtrix to guide the hash learning.
A Hadamard matrix constructed via Sylvester method \cite{Sylvester1867LX} has the following properties
\begin{itemize}
\item It is an $r^*$-order ($r^* = 2^n,n=1,2...$) squared matrix whose elements are either +1 or -1. The coding length $r^*$ of the generated Hadamard matrix is
\begin{equation}
\label{eq2}
r^* = \min\{l|l = 2^n,l\geq r, l\geq k,n=1,2,...\}
\end{equation}
\item  Both its rows and columns are pair-wise orthogonal, which ensures the Hamming distance between any two column vectors is $r^*/2$. Thus, each column of Hada-mard matrix can serve as a hash center satisfying Definition 1.
\end{itemize}
Referring to Eq. (2), there may be  cases when the output code length $r$ does not satisfy: $r = r^*$. To mitigate this problem, Local Sensitive Hashing(LSH) is adopted to transform the hash centers generated by Hadamard matrix to ensure the dimension of centers is consistent with output codes. 
\begin{equation}
\label{eq3}
\tilde{c}_i = sign(\tilde{W}^Tc_i)
\end{equation}
where $\tilde{W} = \{\tilde{w}_i\}_{i=1}^r\in R^{r^*\times r}$ is sampled from the standard Gaussian distribution. The transformed Hadamard matrix  preserves the main properties of the original Hadamard matrix and complies with the requirement of minimal Hamming distance between columns. The detailed theory is developed in \cite{Lin2020Hadamard}\\
\textbf{Semantic hash centers for multi-label data.} Hash centers $\{c_1,c_2,...,c_k\}$ corresponding to $k$ categories respectively have been obtained as described above. For data classifed into two or more categories, the corresponding hash center is the centroid of the multiple centers, each of which relates to a single category.
\subsection{Training Stage}
Through the above process,  we obtain the hash center representation $H^*\in R^{r\times n}$ for all training data in the Hamming space. $H^*$ is also termed as the target codes for the training data. For the $m$-th modality $X^{(m)} = [x_1^{(m)},...,x_n^{(m)}]\in R^{d_m\times n}$ of the training set,  we calculate a nonlinearly  transformed representation $\phi(x^{(m)}_i)=[exp(\frac{\|x_i^{(m)}-a_1^{(m)}\|_F^2}{2\sigma_m^2}),...,exp(\\\frac{\|x_i^{(m)}-a_p^{(m)}\|_F^2}{2\sigma_m^2})]$, where $\{a_j^{(m)}\}_{j=1}^p$ are $p$ anchors that are randomly selected from the $m$-th modality of the training data and $\sigma _m$  denotes the Gaussian kernel parameter. The  $\phi(X^{(m)})\\=[\phi(x^{(m)}_1),...\phi(x^{(m)}_n)]\in R^{p\times n}$ preserve the intra-modality correlation among data within single modality.  The time complexity of this preprocessing phase is $\mathcal{O}(Mnp)$, which is linear to the size of the
training set.

The  heterogenous modalities are projected into a common Hamming space. In this space, data points of the same category are encouraged to migrate towards  a common hash center and those of different categories converge to distinct hash centers. Thus, we have the following
\begin{equation}
\label{eq4}
\min_{W^{(m)}} \sum_{m=1}^M\| H^*-sign(W^{(m)}\phi(X^{(m)}))\|_F
\end{equation}
where $W^{(m)}\in R^{r\times p}$ is the projection matrix of the $m$-th modality, each column of $H^*$ is  the target code of the corresponding training sample, and $\|\cdot\|_F$ denotes the Frobenius norm of a matrix.

In multimedia retrieval, there may potentially be discrepancy between the heterogeneous modalities. Accordingly, it is necessary to gauge the importance of different modalities so as to learn an effective and discriminative hash function. To handle the problem, we transform Eq. (\ref{eq4}) to its equivalent form (see Proof 1):\\
\begin{equation}
\label{eq5}
\begin{split}
\min_{\mu^{(m)},W^{(m)}} &\sum_{m=1}^M\frac{1}{\mu^{(m)}}\| H^*-sign(W^{(m)}\phi(X^{(m)}))\|_F^2\\
&s.t.  \sum_{m=1}^M \mu^{(m)} = 1
\end{split}
\end{equation}
As formulated in Eq.(\ref{eq5}), $\frac{1}{\mu^{(m)}}$  can be considered as a function of  $\mu^{(m)}$.  The more discriminative the $m$-th modality, the  smaller value of $\|H^*-W^{(m)}\phi(X^{(m)})\|_F^2$, and  the larger the corresponding $\frac{1}{\mu^{(m)}}$, and vice versa.\\
\textbf{Proof 1:} Eq.(\ref{eq4}) is equivalent to Eq.(\ref{eq5}). \\
According to the Cauchy-Schwarz inequality, the folllowing (\ref{eq6}) holds.\\
\begin{equation}
\label{eq6}
\begin{split}
&\sum_{m=1}^M\frac{1}{\mu^{(m)}}\| H^*-sign(W^{(m)}\phi(X^{(m)}))\|_F^2\\
&\Leftrightarrow (\sum_{m=1}^M\frac{1}{\mu^{(m)}}\| H^*-sign(W^{(m)}\phi(X^{(m)}))\|_F^2)(\sum_{m=1}^M\mu^{(m)})\\
&\ge (\sum_{m=1}^M\|H^*-sign(W^{(m)}\phi(X^{(m)}))\|_F)^2
\end{split}
\end{equation}
Thus, we can obtain
\begin{equation}
\label{eq7}
\begin{split}
 &(\sum_{m=1}^M\|H^*-sign(W^{(m)}\phi(X^{(m)}))\|_F)^2\\& = \min_{\mu^{(m)}}\sum_{m=1}^M\frac{1}{\mu^{(m)}}\| H^*-sign(W^{(m)}\phi(X^{(m)}))\|_F^2\\
&then \\
&\min_{W^{(m)}}\sum_{m=1}^M\|H^*-sign(W^{(m)}\phi(X^{(m)}))\|_F\\
&\Leftrightarrow \min_{W^{(m)}}(\sum_{m=1}^M\|H^*-sign(W^{(m)}\phi(X^{(m)}))\|_F)^2\\
&\Leftrightarrow \min_{\mu^{(m)},W^{(m)}}\sum_{m=1}^M\frac{1}{\mu^{(m)}}\| H^*-sign(W^{(m)}\phi(X^{(m)}))\|_F^2
\end{split}
\end{equation}

To avoid over-fitting, a regularization term is added to (\ref{eq5}). The overall learning framework the becomes
\begin{equation}
\label{eq8}
\begin{split}
\min_{\mu^{(m)},W^{(m)}} &\sum_{m=1}^M\frac{1}{\mu^{(m)}}\| H^*-sign(W^{(m)}\phi(X^{(m)}))\|_F^2\\
&+ \delta\sum_{m=1}^M\|W^{(m)}\|_F^2\\
&s.t.  \sum_{m=1}^M \mu^{(m)} = 1
\end{split}
\end{equation}
where $\delta$ is a penalty parameter. We relax the objective function to make it tractable computationally, since the sign function makes it difficult to optimize  (\ref{eq8}) directly. The relaxed objective function can be written as:
\begin{equation}
\label{eq9}
\begin{split}
\min_{\mu^{(m)},W^{(m)}} &\sum_{m=1}^M\frac{1}{\mu^{(m)}}\| H^*-W^{(m)}\phi(X^{(m)})\|_F^2
+ \delta\sum_{m=1}^M\|W^{(m)}\|_F^2\\
&s.t.  \sum_{m=1}^M \mu^{(m)} = 1
\end{split}
\end{equation}
We adopt the alternating optimization method  to solve the relaxed problem in (9).
\begin{itemize}
\item Step 1: Update $W^{(m)}$ with other variables fixed. We set the derivative of the objective function with respect to $W^{(m)}(m=1,2....)$  to zero and obtain
\begin{equation}
\label{eq10}
W^{(m)} = \frac{1}{\mu^{(m)}}H^* \phi^T(X^{(m)})(\frac{1}{\mu^{(m)}}\phi(X^{(m)})\phi^T(X^{(m)}) + \delta I)^{-1}
\end{equation}
\item Step 2: Update $\mu^{(m)}$ with other variables fixed.  Gathering the terms relating to $\mu^{(m)}$ , we get the subproblem
\begin{equation}
\label{eq11}
\begin{split}
\min_{\mu^{(m)}\ge 0 }&\sum_{m=1}^M\frac{(G^{(m)})^2}{\mu^{(m)}}   \\ & s.t. \sum_{m=1}^M \mu^{(m)} =1
\end{split}
\end{equation}
where $G^{(m)} = \|H^*-W^{(m)}\phi(X^{(m)})\|_F$.  According to the Cauchy-Schwarz inequality,  the optimal $\mu^{(m)}$ can be obtained as
\begin{equation}
\label{eq12}
\mu^{(m)}= \frac{G^{(m)}}{\sum_{m=1}^MG^{(m)}}
\end{equation}
\end{itemize}
After obtaining the optimal parameters, we use the hash function $f= sgn(\sum_{m=1}^M\frac{1}{\mu^{(m)}}W^{(m)}\phi(X_q^{(m)}))$ that can fuse multiple modal information to generate the unified representation of multi-modal data.
\subsection{Hash Encoding with Dynamic Weights}
Unfortunately, the fixed weights learned from Eq. (\ref{eq12}) can not capture the variations of dynamic data in the process of hash coding. Thus, the weights should be adjusted dynamically for each specific instance content. Motivated by this intuition, we adopt a self-weighting scheme in form of online training to obtain more accurate hash codes for newly arriving  multimedia data. In the encoding stage, we assume that data appear in the manner of the data stream. Newly arriving data will be hash coded and archived in the database. The adaptive online hash encoding process is given as
\begin{equation}
\label{eq13}
\begin{split}
\min_{B_q,\mu_q^{(m)}}&\sum_{m=1}^M\frac{1}{\mu_q^{(m)}}\|B_q-W^{(m)}\phi(X_q^{(m)})\|_F^2
\\ & s.t. \sum_{m=1}^M \mu^{(m)} =1, B_q\in \{-1,1\}^{r\times n_q}
\end{split}
\end{equation}
where $B_q$ and $n_q$ are the hash codes and the number of the new instances respectively. $\phi(X_q^{(m)})$ is the nonlinearly transformed representation of the $m$-th modality of the newly coming instances.\\
We solve the problem in Eq. (\ref{eq13}) by alternative updating the following variables iteratively.\\
Update $\mu_q^{(m)}$ with fixed $B_q$. The optimal solution of $\mu_q^{(m)}$ is obtained as
\begin{equation}
\label{eq14}
\mu_q^{(m)}= \frac{G_q^{(m)}}{\sum_{m=1}^MG_q^{(m)}}
\end{equation}
where $G_q^{(m)} = \|B_q-W^{(m)}\phi(X_q^{(m)})\|_F$.\\
Update $B_q$ with fixed $\mu_q^{(m)}$.  We can get a closed solution as
\begin{equation}
\label{eq15}
B_q = sgn(\sum_{m=1}^M\frac{1}{\mu_q^{(m)}}W^{(m)}\phi(X_q^{(m)}))
\end{equation}
The optimal $B_q$ is viewed as the final output codes by above optimization process. The proposed Adaptive Multi-modal Fusion Hashing (AMFH) is summarized in Algorithm \ref{alg1}. For the missing modality problem in which the partial modalities is unknown, the corresponding weighting factor is set zero. In the experiment section, we take image modality and text modality as an example to present some discussions. 

\renewcommand{\algorithmicrequire}{\textbf{Input:}}
\renewcommand{\algorithmicensure}{\textbf{Output:}}
\begin{algorithm}
\caption{Adaptive Multi-modal Fusion Hashing}
Training stage:
\begin{algorithmic}[1]
\REQUIRE Training set $X^{(m)} = [x_1^{(m)},...,x_n^{(m)}]\in R^{d_m\times n},(m=1,...,M)$ with category information.\\
Generating target codes $H^*$\\
Calculate the transformed representation $\phi(X^{(m)})=[\phi(x^{(m)}_1),...\phi(x^{(m)}_n)]\in R^{p\times n}$\\
Initialize $W^{(m)}(m=1,...,M)$ and $\mu^{(m)}(m=1,...,M)$\\
\REPEAT
\STATE  Update $W^{(m)}(m=1,...,M)$ according to (\ref{eq10});\\
   \STATE  Update $\mu^{(m)}(m=1,...,M)$ according to (\ref{eq11});
\UNTIL convergence
\ENSURE $W^{(m)}(m=1,...,M)$

Hash encoding stage:\\
\textbf{for q =1,...T do}\\
\quad\quad Receive newly arriving $X_q$\\
\quad\quad\textbf{repeat}\\
\quad\quad\quad Update $\mu^{(m)}(m=1,...,M)$ according to (\ref{eq14});\\
\quad\quad\quad Update $B_q$ according to (\ref{eq15});\\
\quad\quad\textbf{until} convergence\\
\quad\quad\textbf{Output} $B_q$\\
\textbf{end for} \\
\end{algorithmic}
\label{alg1}
\end{algorithm}

\begin{figure*}[!t]
\centering
\includegraphics[trim={0mm 60mm 0mm 0mm},clip,width=\textwidth]{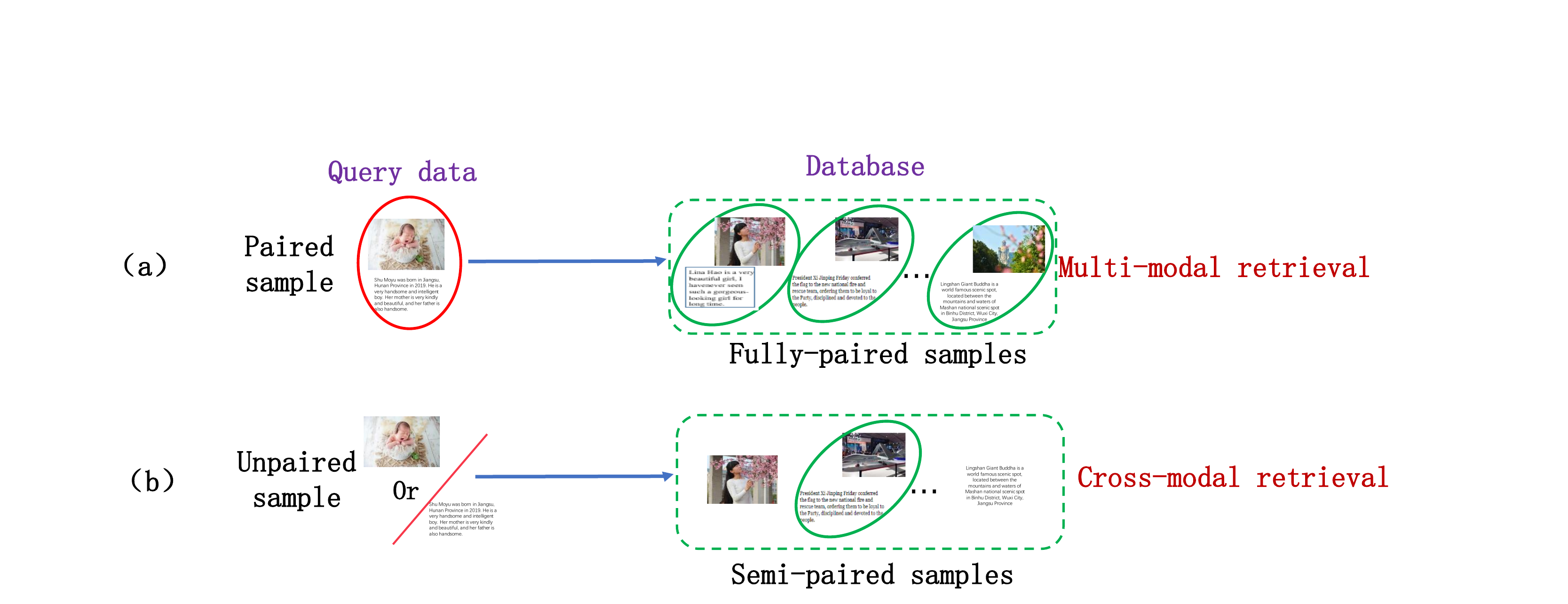}
\caption{Two data scenarios in hash encoding process}
\label{figcase}
\end{figure*}

\begin{table*}[t]
\caption{mAP Comparison of multi-modal retrieval for different bits}\smallskip
\centering
\smallskip\begin{tabular}{c|cccc|cccc|cccc}
\hline
\multirow{2}{*}{Methods} & \multicolumn{4}{c|}{WiKi} &  \multicolumn{4}{c|}{Pascal VOC 2007}  &  \multicolumn{4}{c}{NUS-WIDE} \\
\cline{2 - 13}
&16&32&64&128&16&32&64&128&16&32&64&128\\
\hline
ITQ\cite{gong2013iterative}&0.5122&0.5359&0.5490&0.5532&0.7586& 0.7975&0.8053&0.8061&0.3724&0.3751&0.3776&0.3789\\
\hline
LSH\cite{datar2004locality-sensitive}&0.4306&0.4712&0.5085&0.5276&0.4402&0.5591&0.6628&0.7262&0.3421&0.3554&0.3544&0.3672\\
\hline
DLLE\cite{ji2017toward}&0.5234&0.5330&0.5466&0.5506&0.7629&0.8068&0.8131&0.8193&0.3738&0.3782&0.3794&0.3823\\
\hline
HCOH\cite{lin2018supervised}&0.5450&0.5474&0.5494&0.5490&0.2436&0.6050&0.6070&0.6072&0.3232&0.3451&0.3434&0.3645\\
\hline
MFH\cite{Song2013Effective}&0.4630&0.5040&0.5455&0.5569&0.5364&0.6376&0.6941&0.7216&0.3673&0.3752&0.3803&0.3815\\
\hline
MVLH\cite{Shen2015Multiview}&0.3027&0.3166&0.3000&0.3045&0.5469&0.6324&0.6995&0.7203&0.3363&0.3339&0.3324&0.3284\\
\hline
FOMH\cite{lu2019flexible}&0.4408&0.4813&0.5092&0.5149&0.4436&0.5676&0.6504&0.7343&0.3863&0.4450&0.5244&0.5519\\
\hline
OMH-DQ\cite{Lu2019Online}&0.4117&0.4393&0.4556&0.4319&0.5673&0.7040&0.8096&0.8542&0.5223&0.5381&0.5823&0.5957\\
\hline
EPAMH\cite{zheng2020efficient}&-&-&-&-&0.7704&0.8193&0.8493&0.8738&0.4025&0.3969&0.3915&0.3879\\
\hline
FDCMH\cite{zheng2019fast}&0.5986&0.6392&0.6409&0.6531&\bf{0.8305}&0.8517&0.8727&0.8921&0.5632&0.5820&0.6018&0.6201\\
\hline
SIDMH\cite{lu2020semantic}&0.5666&0.6219&0.6307&0.6416&0.8138&0.8346&0.8511&0.8634&0.5828&0.5976&0.6055&0.6120\\
\hline
Ours&\begin{bfseries}0.6580\end{bfseries}&\begin{bfseries}0.6674\end{bfseries}&\begin{bfseries}0.6677\end{bfseries}&\begin{bfseries}0.6752\end{bfseries}&0.7715&\begin{bfseries}0.9001\end{bfseries}&\begin{bfseries}0.9009\end{bfseries}&\begin{bfseries}0.9016\end{bfseries}&\begin{bfseries}0.6190\end{bfseries}&\begin{bfseries}0.6240\end{bfseries}&\begin{bfseries}0.6271\end{bfseries}&\begin{bfseries}0.6385\end{bfseries}\\
\hline
\end{tabular}
\label{table1}
\end{table*}

\section{Experiment}
\label{exeriments}
In this section, we conduct multi-modal retrieval experiments on three widely-used multi-modal datasets to verify the performance of the proposed method. Some extended experiment are also carried out to investigate  the cross-modal retrieval. Figure \ref{figcase} shows two possible data scenarios. For cross-modal retrieval scenario, the paired relationship is not satisfied strictly and query data is single modality. In our experiments, $p$ is set to 1000. Our experiments are executed on a Windows 10 platform based desktop machine with 12GB memory and 4-core 3.6GHz CPU.
\subsection{Datasets}
\label{dataset}
\textbf{WiKi} \cite{Rasiwasia2010ANewApproach} is a multi-modal  single-label dataset which consists of 2866 multimedia documents of 10 categories. We directly generate one hash center for each category. Each image is represented by 128-dimensional SIFT histogram vector, while text is represented as a 10-dimensional feature vector generated by latent Dirichlet allocation. A random subset of 2173 multimedia samples is used as the offline training set and the retrieval set and the remaining 963 samples as the query set.  \\
\textbf{Pascal VOC 2007} \cite{wei2017cross-modal} contains 9963 images of 20 categories. Each image and  associated 399 tags with the image compose a multimedia sample. In this dataset, We employ the 4096-dimensional CNN feature to represent the visual object and the 798 dimensional tag ranking feature is employed as the text feature. A random subset of 2000 samples is provided as the offline training set, and the remaining samples are divided into query set and retrieval set, containing 963 and 7000 samples respectively. \\
\textbf{NUS-WIDE} \cite{Chua2009NUSWIDE} is comprised of 269648 multi-modal samples of 81 concepts. In our experiments, we only keep 186577 samples of  the top ten most frequent concepts. The image modality is represented by a 500 dimensional bag- of-visual words and the 1000 dimensional tag occurrence vector is employed as text modality feature. A random subset of 1866 samples for query set and 184711 samples for retrieval set. 5000 samples are randomly selected from the retrieval set for the offline training stage.

Pascal VOC 2007 and NUS-WIDE are two multi-lable datasets. For those multimedia samples with multiple labels, we first generate hash centers for single category, then calculate the centroid of the multi-centers as the semantic target codes of this sample.

\subsection{Baselines and Evaluation Scheme}
Seven related and state-of-the-art hashing methods are adopted for comparison in multi-modal retrieval. These baselines can be divided into two categories. (1) Multi-modal hashing methods including MFH \cite{Song2013Effective}, MVLH \cite{Shen2015Multiview}, FOMH\\\cite{lu2019flexible}, OMH-DQ \cite{Lu2019Online}, EPAMH\cite{zheng2020efficient}, FDCMH\cite{zheng2019fast} and SIDMH\\\cite{lu2020semantic}; (2) Single-modal hashing methods including ITQ \cite{gong2013iterative}, LSH \cite{datar2004locality-sensitive}, DLLE \cite{ji2017toward}, HCOH \cite{lin2018supervised}. Since the single-modal methods can not deal with multiple modalities simultaneously, we concatenate multiple modalities as the input feature for a fair comparison. It is noted that EPAMH is only suitable for this case where the dimension of original multimodal feature larger the length of hash codes. On WiKi dataset, EPAMH is not compared in experiments since the textual dimension is less than 16. For fair comparison, the text features and image features descripted in section \ref{dataset} are imported into the deep module in the SIDMH framework to replace its feature input. For Additionally, we compare our method with the following six supervised cross-modal hashing methods : MDBE\cite{wang2016multimodal}, SePH\cite{lin2016cross}, DCH\cite{xu2017learning}, DLFH\cite{jiang2019discrete}, KDLFH\cite{jiang2019discrete}, EDMH\cite{chen2020enhanced}. We adjust the parameters of each method to take values from the candidate range given in the original papers and report the best results. The performance is evaluated by Mean Average Precision (mAP) \cite{Yi2012Co,Zhang2014Large-Scale}. For a query $q$, the Average Precision (AP) is defined as follows
\begin{equation}
\label{eq16}
AP(q) =\frac{1}{l_q}\sum_{m=1}^RP_q(m)\delta_q(m)
\end{equation}
where $P_q(m)$ denotes the accuracy of the top $m$ retrieval results; $\delta_q(m)=1$ if the $m$-th position is the true neighbour of the query $q$, and otherwise $\delta_q(m)=0$; $l_q$ is the correct statistics of top $R$ retrieval results. The mAP is defined as the mean of the average precisions of all the queries.
\subsection{Accuracy Comparison}
The experimental results  on WiKi, Pascal VOC 2007 and NUS-WIDE are presented in Table \ref{table1}. We can clearly observe that our method consistently outperforms all the baselines used in the comparison, when the code length varies from 16 bits to 128 bits. As the code length increases, the performance of our method improves slightly. This behaviour demonstrates that our method is not sensitive to the code length and can achieve satisfactory performance even with short codes. AMFH is superior to single modal hashing methods with different hash code lengths. The reason for better performance of AMFH is that it can reduce the redundant information among multiple modalites. AMFH achieves much higher mAP scores than MFH and MVLH, for the reason that the adaptive encode strategy can improve the quality of hash codes. Compared with OMH-DQ, our method achieves an average improvement of 23\%, 13\% and 7\% on WiKi, Pascal VOC 2007 and NUS-WIDE respectively. This indicates that our method which assigns the hash centers for those samples with different category information via Hadamard matrix can generate effective hash codes in large-scale applications. FOMH and FDCMH are two classic asymmetric hashing models, which aims to learn optimal binary codes for each category. In Table \ref{table1}, we can observe that the accuracy of our proposed AMFH is competitive compared with FOMH and FDCMH. Specifically, The average score of AMFH is higher than FDCMH by 2.5\%. SIDMH ultilizes class label and a hierarchical model to guide the generation of hash codes. Although the complementary information between multimodal data can be well mined in SIDMH, the discrete solution in optimization is only the approximation of the optimal solution. Our method obtain better experimental results than SIDMH on three datasets. The comparison results imply that the customized hash center in this article is very discriminative, so that different categories of data can be effectively separated in the Hamming space.

\begin{figure*}[t]
\centering
\subfigure[WiKi]{
\includegraphics[width=.3\textwidth]{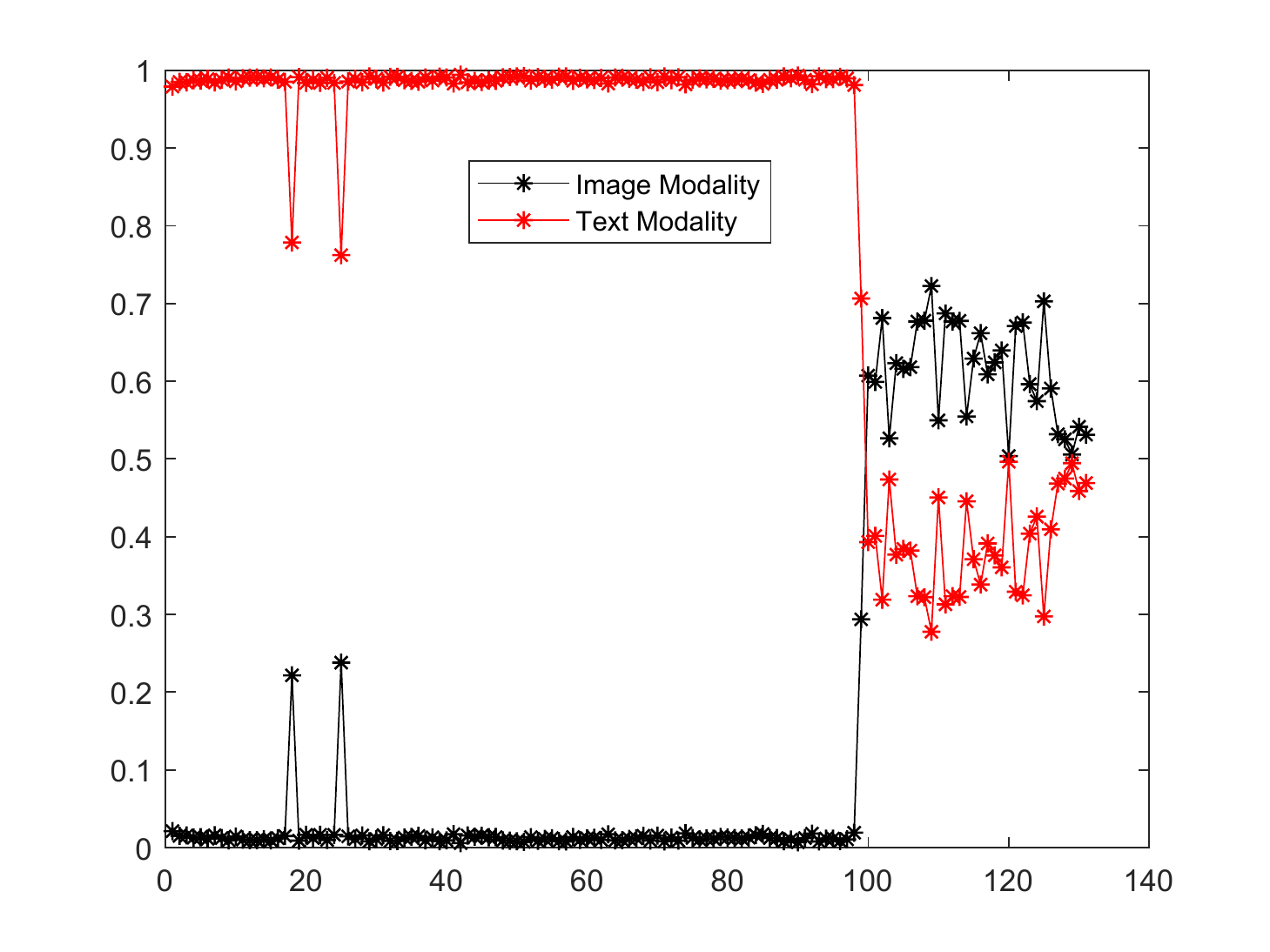}}
\subfigure[Pascal VOC 2007]{
\includegraphics[width=.3\textwidth]{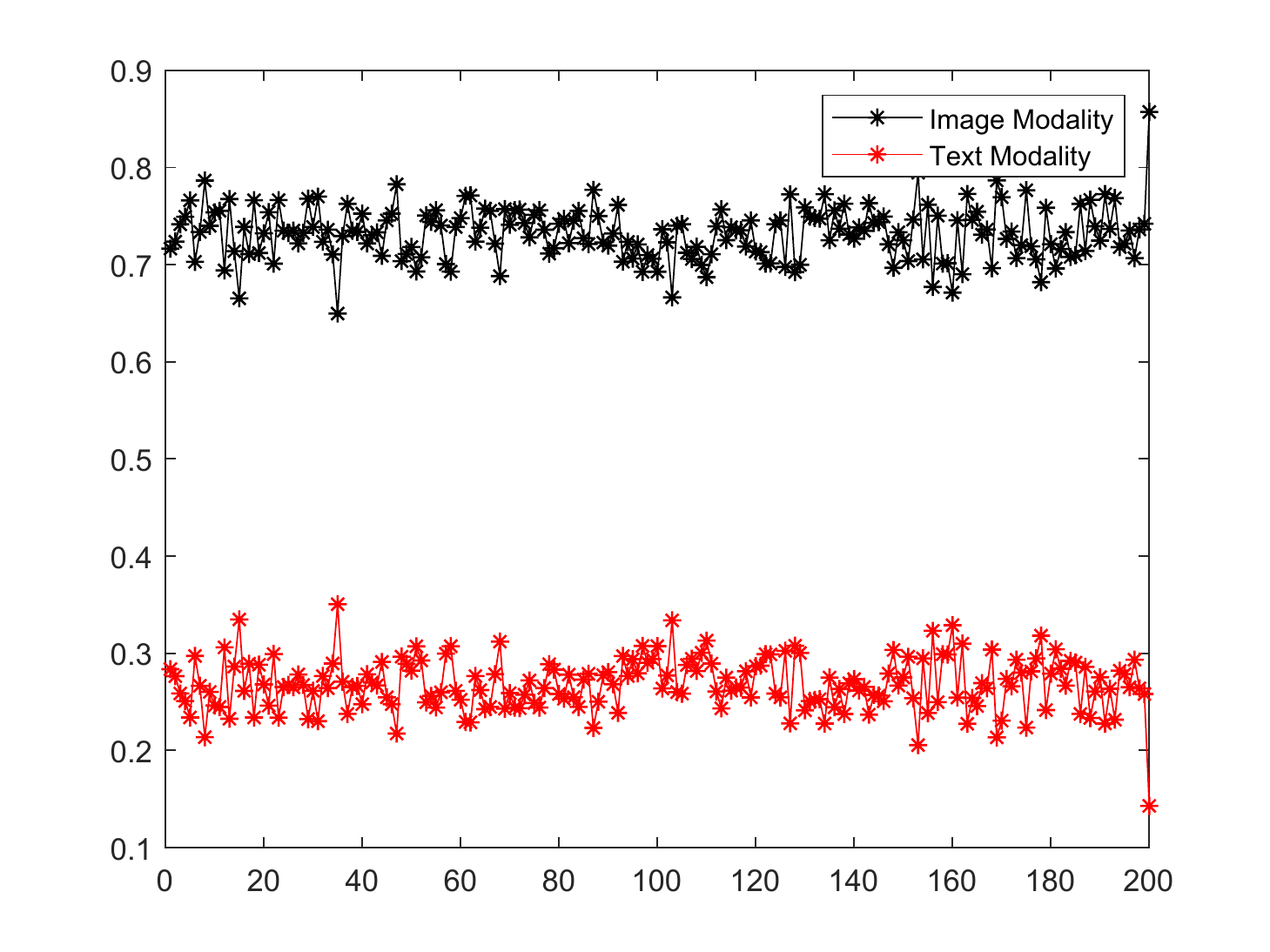}}
\subfigure[NUS-WIDE]{
\includegraphics[width=.3\textwidth]{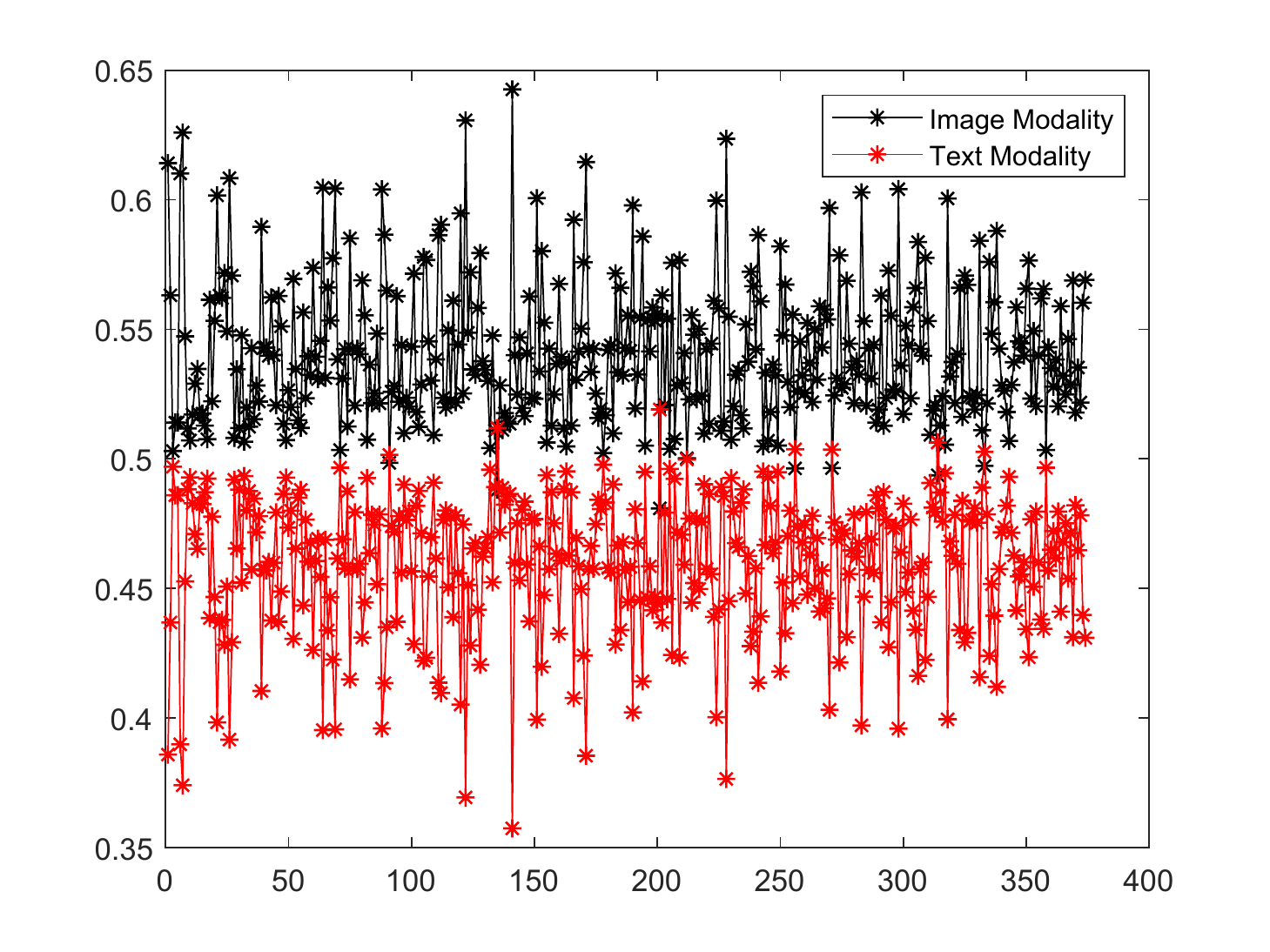}}
\caption{Visualization of modality weights adapted to dynamic data}
\label{fig4}
\end{figure*}

\begin{figure}[t]
\centering
\includegraphics[width=\columnwidth]{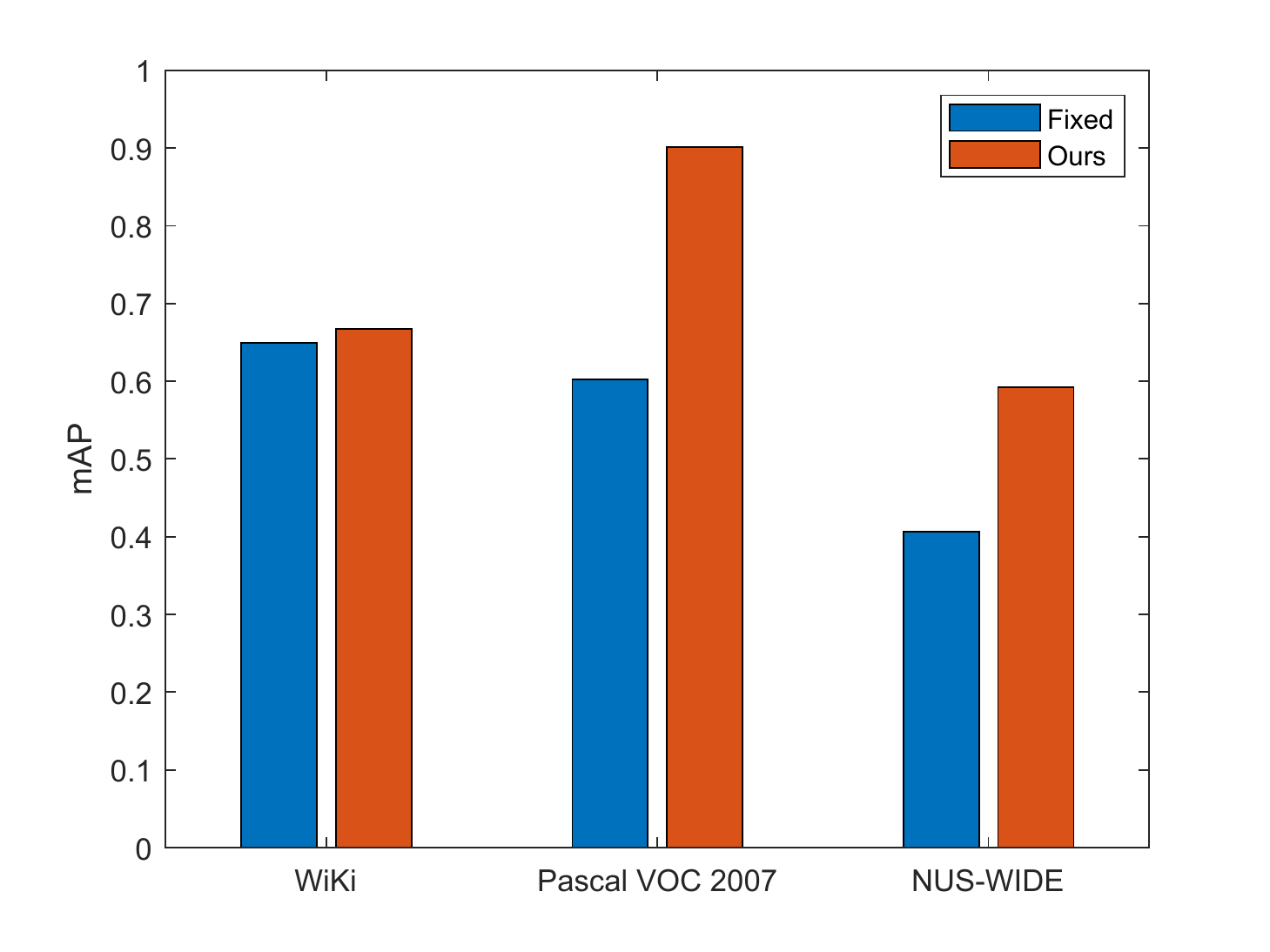} 
\caption{Results of the ablation experiments  on the adaptive online stage}
\label{fig3}
\end{figure}

\begin{table}[t]
\caption{Comparison of training  time (seconds)}\smallskip
\centering
\smallskip\begin{tabular}{c|ccc}
\hline
\multirow{2}{*}{Methods} & \multicolumn{3}{c}{Training time (s)} \\
\cline{2 - 4}
&WiKi&Pascal VOC 2007&NUS-WIDE\\
\hline
ITQ\cite{gong2013iterative}&0.5033&87.7270&2.5625\\
\hline
LSH\cite{datar2004locality-sensitive}&0.0129&2.9982&0.1014\\
\hline
DLLE\cite{ji2017toward}&139.1312&147.1619&1461.2236\\
\hline
HCOH\cite{lin2018supervised}&0.2302&9.9715&3.1992\\
\hline
MFH\cite{Song2013Effective}&2.7651&25.9216&19.0934\\
\hline
MVLH\cite{Shen2015Multiview}&184.7249&452.2433&913.1369\\
\hline
FOMH\cite{lu2019flexible}&1.2867&41.4450&3.1061\\
\hline
OMH-DQ\cite{Lu2019Online}&8.3657&192.0446&70.1140\\
\hline
EPAMH\cite{zheng2020efficient}&-&94.1752&1.9401\\
\hline
FDCMH\cite{zheng2019fast}&73.4721&99.6400&165.2437\\
\hline
SIDMH\cite{lu2020semantic}&43.15121&204.3702&133.4563\\
\hline
Ours&0.3724&23.8603&1.6507\\
\hline
\end{tabular}
\label{table2}
\end{table}

\begin{figure}[t]
\centering
\includegraphics[width=0.9\columnwidth]{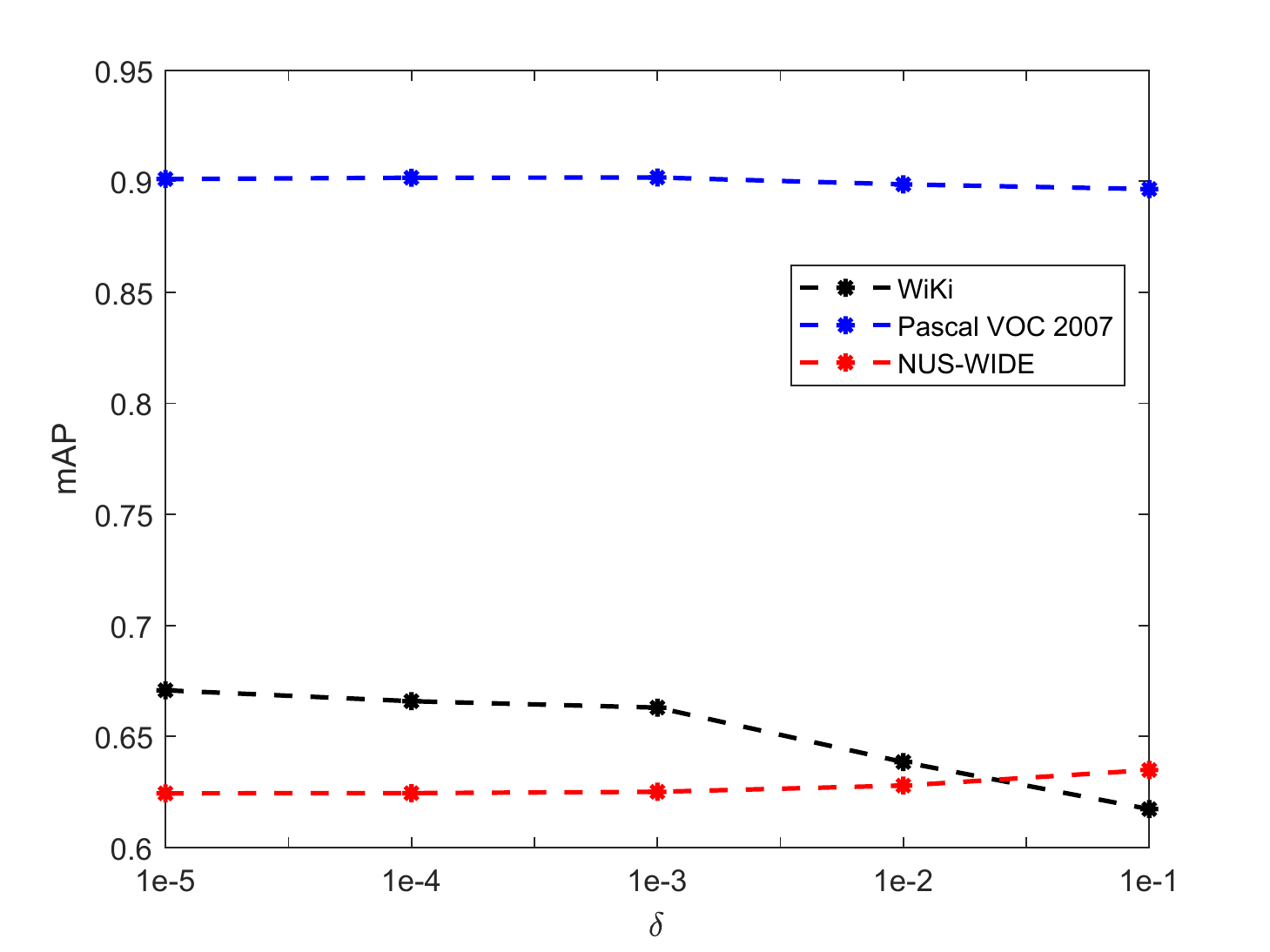}
\caption{Performance variation with respect to $\delta$. }
\label{fig2}
\end{figure}

\begin{figure*}[htbp]
\centering
\subfigure[WiKi]{
\includegraphics[width=.3\textwidth]{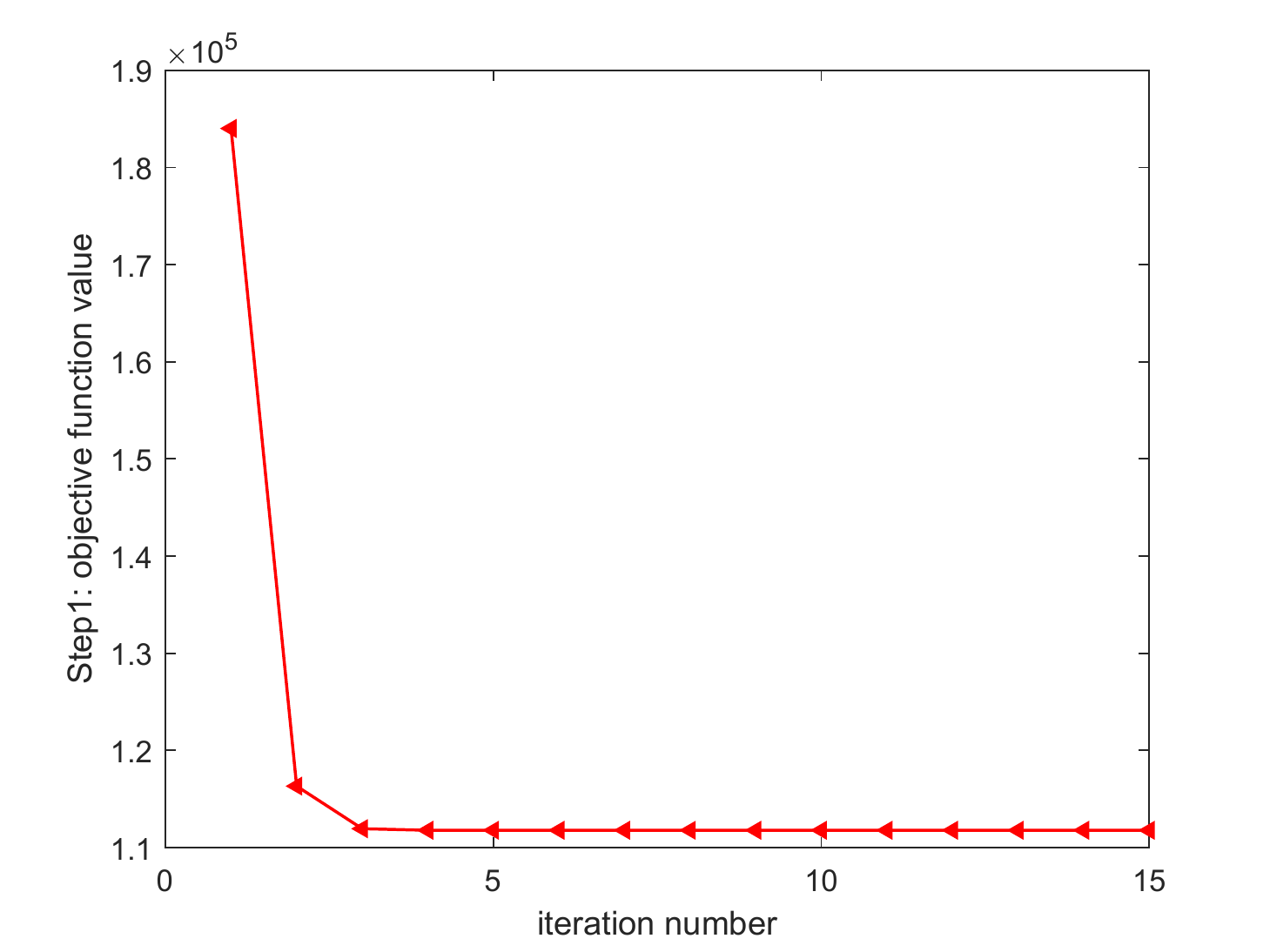}}
\subfigure[Pascal VOC 2007]{
\includegraphics[width=.3\textwidth]{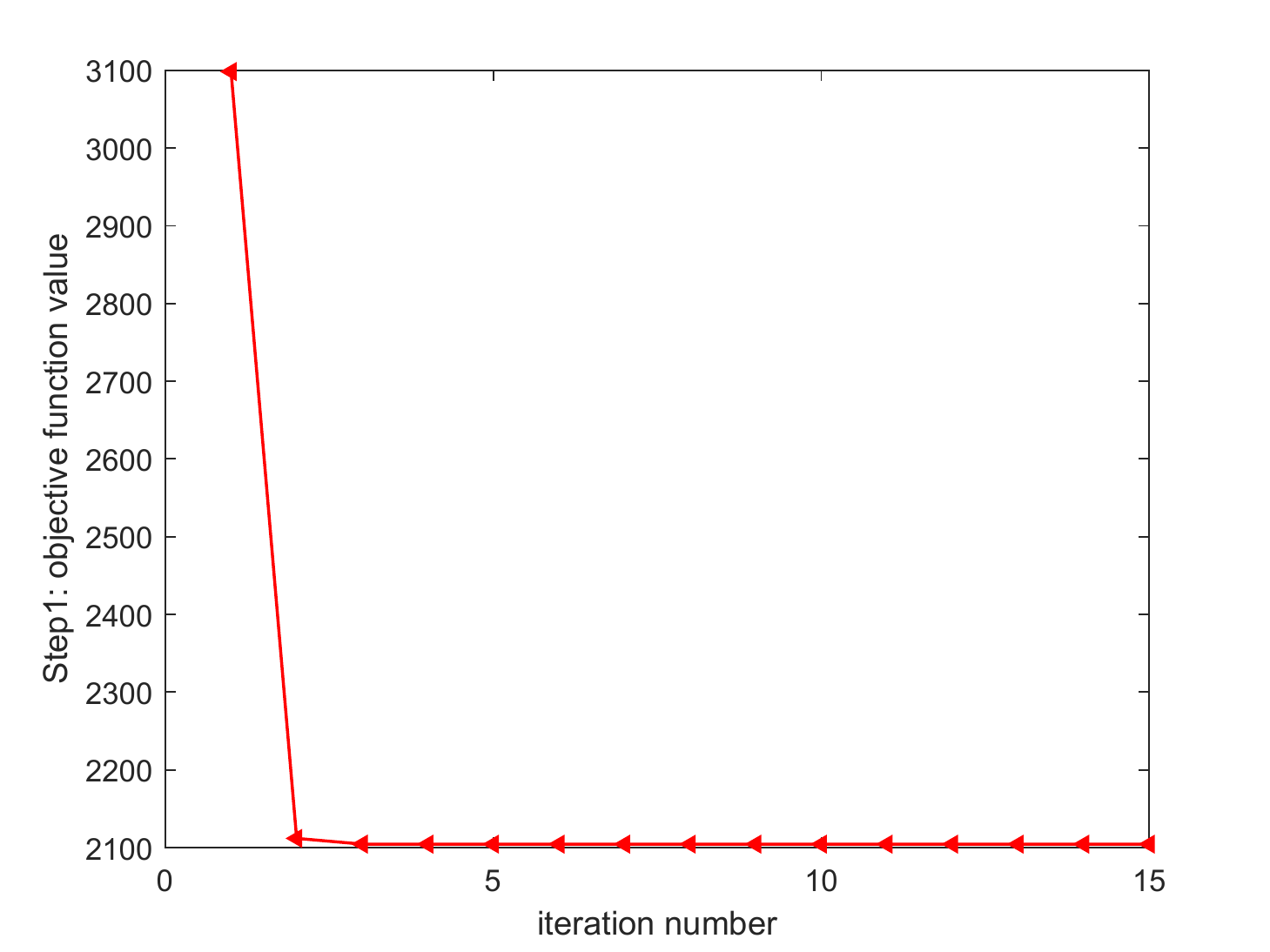}}
\subfigure[NUS-WIDE]{
\includegraphics[width=.3\textwidth]{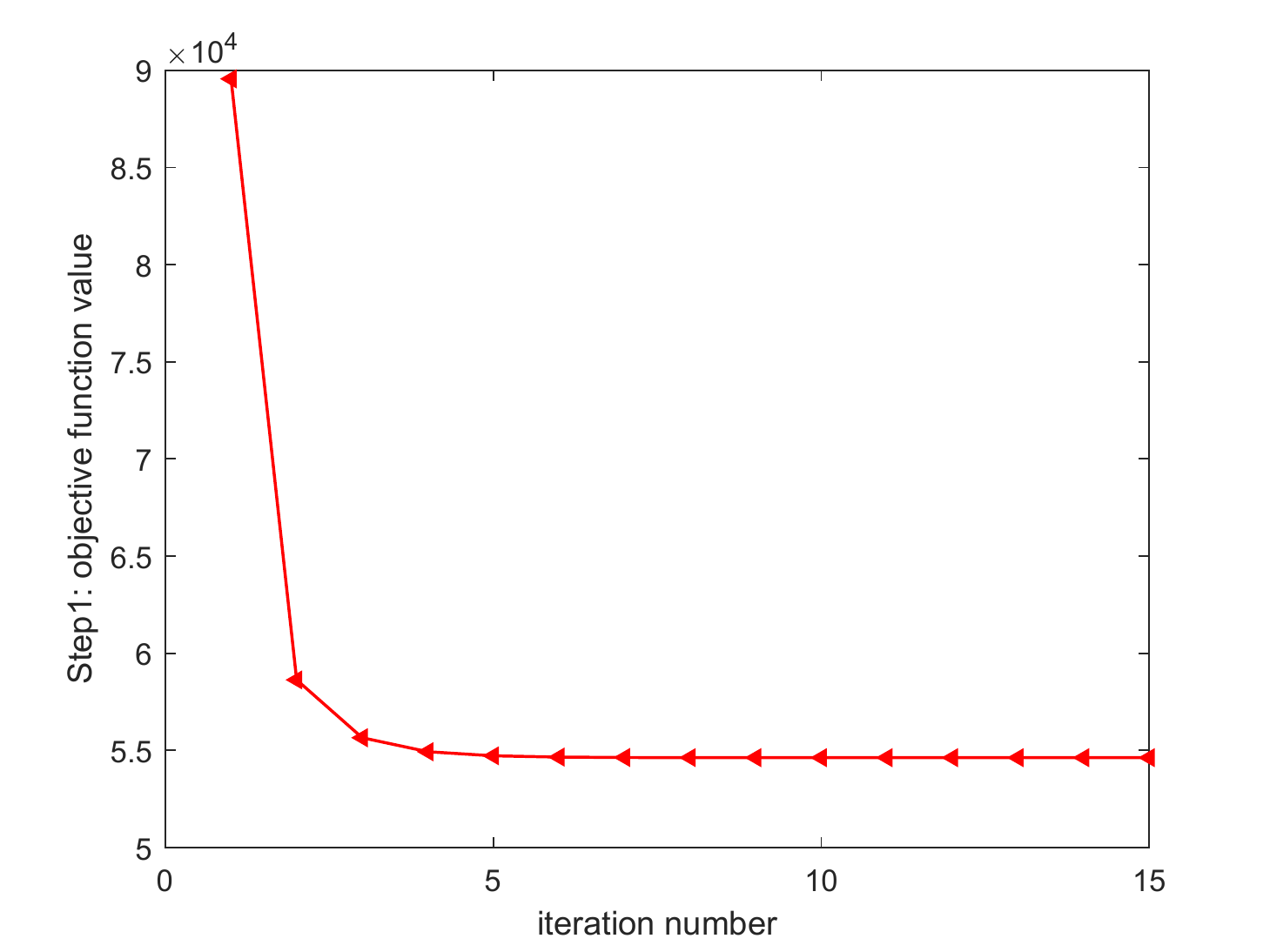}}
\caption{Convergence curves on  Wiki (a), Pascal VOC 2007 (b) and NUS-WIDE (c).}
\label{fig5}
\end{figure*}

\textbf{Ablation study} In our method, the projection matrices are learned during the training stage and the hash codes of newly coming multimedia data are generated in an online learning mode in order to capture the variations inherent in the data. In order to validate the effectiveness of the proposed adaptive online strategy, we conduct ablation experiments. Let 'fixed' indicate the weights learned in the training stage. They are applied to generate hash codes for newly coming data. As shown in Figure \ref{fig3}, our method exhibits a considerable improvement over 'fixed'.  Figure \ref{fig4} shows the dynamic variation of the weight of each modality for a new data batch. Note that the variations for  Pascal VOC 2007 and NUS-WIDE are larger than for WiKi. From Figure \ref{fig3} and Figure \ref{fig4}, it is apparent that the improvement gained in Precision is correlated with the extent of variations of the dynamic weight. This implies that the weight adaptation is very important. It impacts on the  performance beneficially, especially for data with large diversity.
\subsection{Run Time Comparison}
In this subsection, we investigate the training time of the proposed method and compare it with baselines by conducting experiments on WiKi, Pascal VOC 2007 and NUS-WIDE datasets respectively. The statistics of the results are reported in Table \ref{table2}. LSH is a popular data-independent method. It is obvious that the computation cost of LSH is relatively low. HCOH is also a supervised method based on Hadamard matrix, but its optimization does not involve the matrix inverse operation. Except for LSH and HCOH,  our method is faster than the other methods compared. Although the training time of our method is slightly slower than that of LSH and HCOH, its performance is much better. Since there is no need to learn the hash matrix in AMFH, this improves the training efficiency of the model. It is obvious that our method requires least training time than OMH-DQ, FOMH, FDCMH and SIDMH, which shows that AMFH greatly reduces the computation cost with higher accuracy. 

\subsection{Parameter sensitivity and Convergence analysis}
There is only one penalty parameter $\delta$ to avoid overfitting in our model. In order to explore its effect on the performance of our model, we vary  its values in the range of $\{1e^{-5}, 1e^{-4},1e^{-3},1e^{-2},1e^{-1}\}$. The performance curves are plotted in Figure \ref{fig2}. We can see that a degradation commences on WiKi from 1e-3. In contrast, the performance is relatively stable for a large range of values on Pascal VOC 2007 and NUS-WIDE, which may be because the overfitting is less likely to happen on larger datasets. In conclusion, our model is insensitive to the parameter and can flexibly be applied, especially to larger-scale multimedia retrieval problems.

\begin{figure*}[!t]
\centering
\subfigure[Image query Text]{
\includegraphics[width=.45\textwidth]{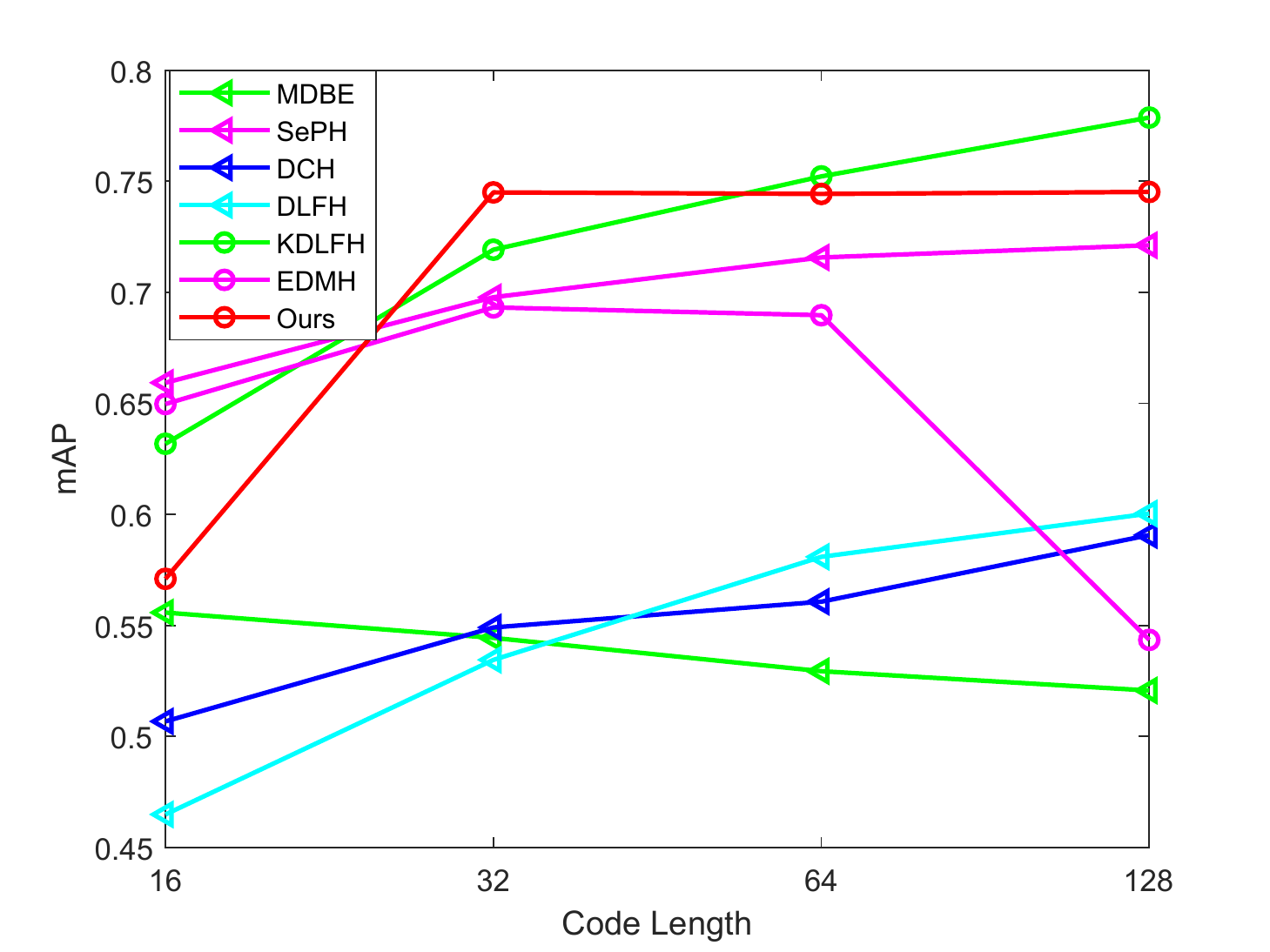}}
\subfigure[Text query Image]{
\includegraphics[width=.45\textwidth]{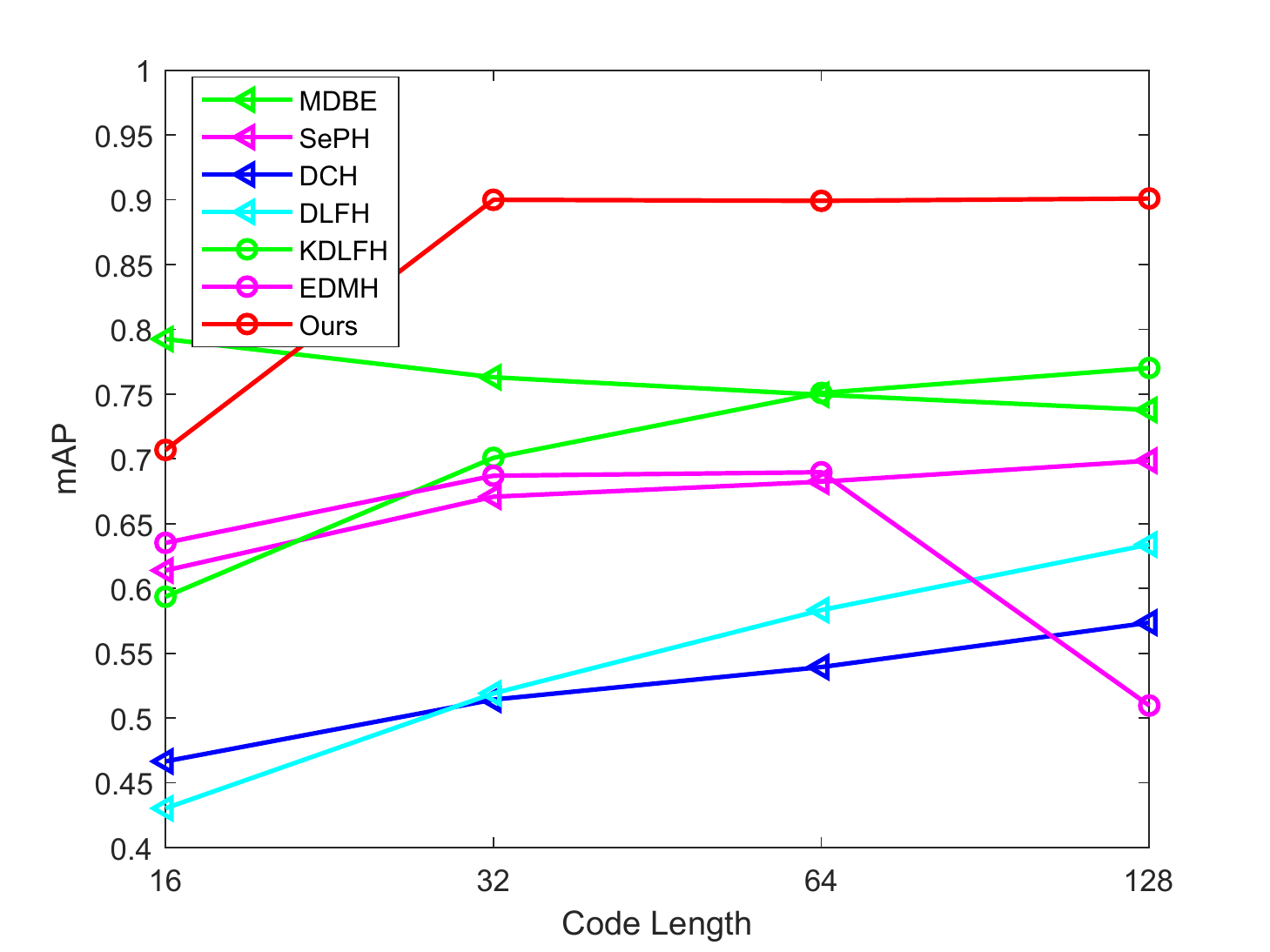}}
\caption{The comparison results for cross-modal retrieval  on Pascal VOC 2007, (a) Image query Text, (b) Text query Image}
\label{fig6}
\end{figure*}

\begin{figure*}[!t]
\centering
\subfigure[Image query Text]{
\includegraphics[width=.45\textwidth]{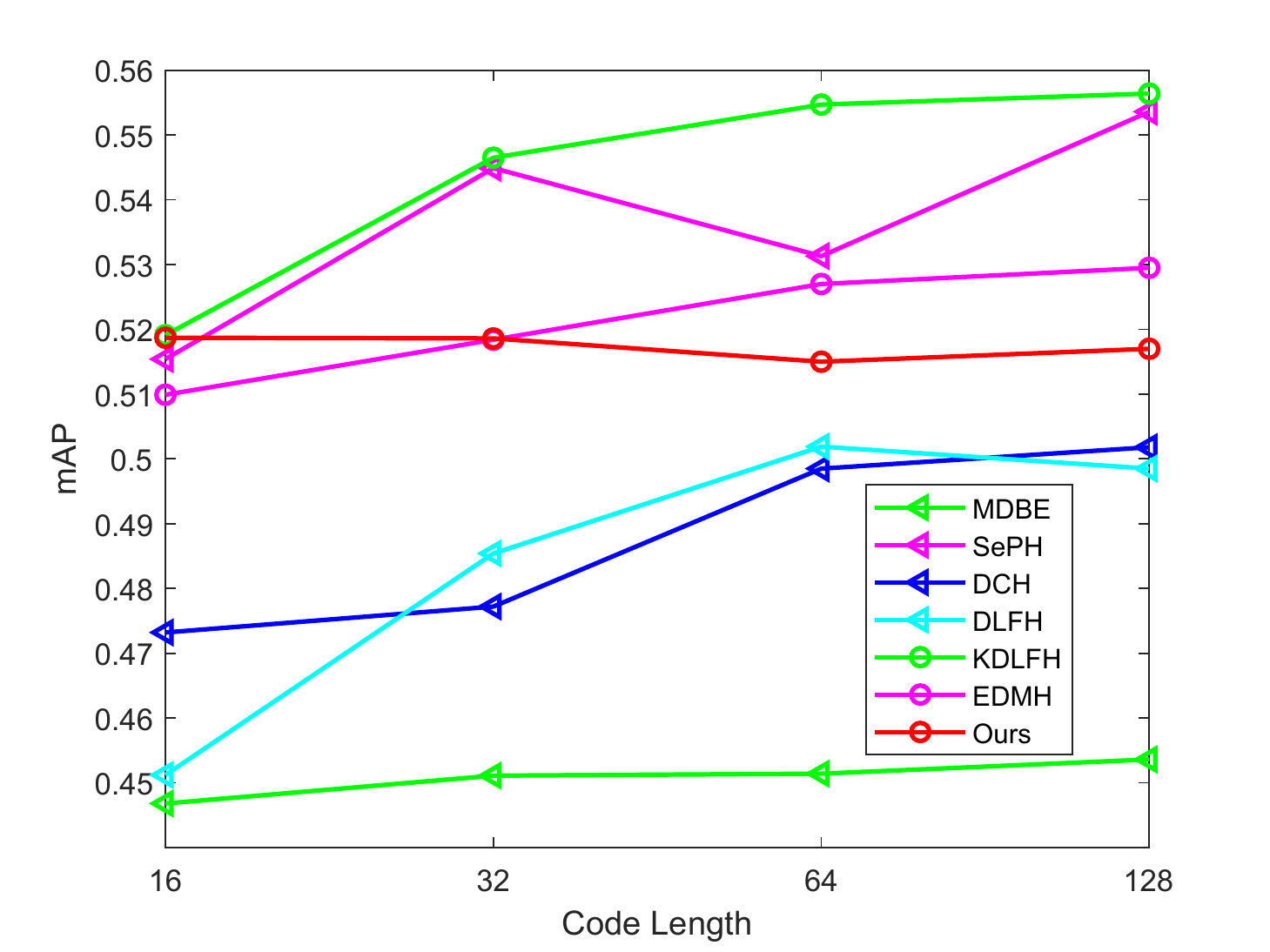}}
\subfigure[Text query Image]{
\includegraphics[width=.45\textwidth]{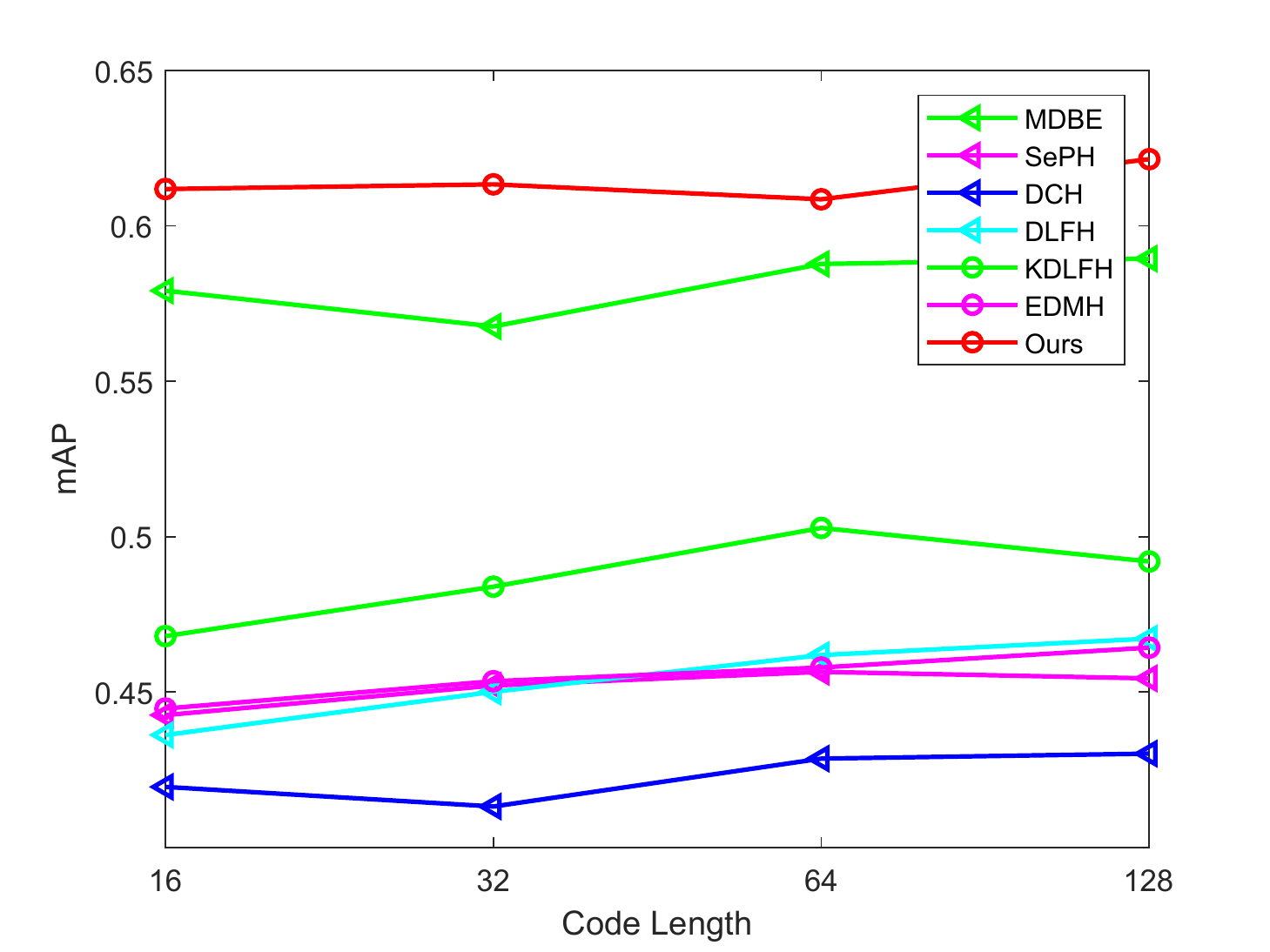}}
\caption{The comparison results for cross-modal retrieval  on NUS-WIDE, (a) Image query Text, (b) Text query Image}
\label{fig7}
\end{figure*}

\begin{table*}[!t]
\caption{The cross-modal retrieval results under the different proportion setting of paired samples}\smallskip
\centering
\smallskip\begin{tabular}{ccccccc}
\hline
\multirow{2}{*}{Dataset} & \multirow{2}{*}{Tasks} &\multicolumn{5}{c}{The proportion of paired samples } \\
\cline{3 - 7}
&&0.1&0.3&0.5&0.7&0.9\\
\cline{1-7}
\multirow{2}{*}{Pascal VOC 2007} &Image query Text&0.7437&0.7442&0.7446&0.7449&0.7450\\
&Text query Image&0.7391&0.7791&0.8181&0.8516&0.8858\\
\multirow{2}{*}{NUS-WIDE} &Image query Text&0.5029&0.5058&0.5079&0.5103&0.5125\\
&Text query Image&0.4726&0.5139&0.5450&0.5770&0.6059\\
\hline
\end{tabular}
\label{table3}
\end{table*}

The optimisation process based on the updating rule (see Eq.(\ref{eq10}) and Eq. (\ref{eq12})) is decreasing the objective function monotonically, and rapidly converges to the minimum. This is shown by the results of the experiments on WiKi, Pascal VOC 2007 and NUS-WIDE using our model with the code of 128-bit length. The convergence curves obtained on the three datasets are plotted in Figure \ref{fig5}. (The convergence trend for the hash codes of other length is similar.) As shown in Figure \ref{fig5}, our model converges within 5 iterations on WiKi, Pascal VOC 2007 and NUS-WIDE.

\subsection{Comparison experiments on Cross-modal retrieval}
In this section, we explore the performance of our model for the data scenario with missing modalities. A single modality as a query sample to retrieve another modal data from the database. As seen in Figure \ref{figcase} (b), the partial modality of some instances is missing. We need to encode these missing data individually, while the fully-paired multimodal data is still represented by the unified hashing features. The cross-modal retrieval tasks including Image query Text and Text query Image are performed on Pascal VOC 2007 and NUS-WIDE. Figure \ref{fig6} and Figure \ref{fig7} show the comparative experimental results on  Pascal VOC 2007 and NUS-WIDE respectively. From the retrieval results, we can observe that our proposed method achieves comparable performance to the best baseline when the hash lengths are set to 64 bit and 16 bit on Pascal VOC 2007 and NUS-WIDE respectively. On the Text query Image task, AMFH achieves much higher mAP scores than all comparison methods. Note that the database is assumed as fully-paired multi-modal data while the query is considered as the unpaired ones in Figure \ref{fig6} and Figure \ref{fig7}. The performance on the Text query Image task is superior to that on the Image query Text task. Further, we set the proportion of paired samples in the database as 0.1, 0.3, 0.5, 0.7, and 0.9 respectively to observe the performance variation. The experimental results on Pascal VOC 2007 and NUS-WIDE are recorded in Table \ref{table3}. It is easy to find that the variation performance on the Image query Text task is smaller than that on the Text query Image task with the increase of the proportion of paired samples in the database. The possible reason is that text modality is much better to represent the high-level semantic content of the corresponding object than image modality. 

\section{Conclusion}
\label{conclusion}
In this paper, we proposed a novel multi-modal hashing method where Hadamard matrix is introduced to generate a discriminative hash center for each content category. Our model exhibits strong discriminative capability and is computationally light. As it is not highly sensitive to hyperparameters, it can be applied  very flexibly. The results of the experiments conducted on several public multi-modal datasets demonstrate the superior accuracy and efficiency of the proposed method, as compared to the state-of-the-art algorithms.

\ifCLASSOPTIONcaptionsoff
  \newpage
\fi



%
\bibliographystyle{IEEEtran}
\bibliography{my}

%








\end{document}